\documentclass[12pt]{article}
\usepackage[a4paper, margin=1in]{geometry}
\usepackage{graphicx}
\usepackage{amsmath, amssymb}
\usepackage{caption}
\usepackage{authblk}
\usepackage{hyperref}
\usepackage{titlesec}
\usepackage{float}
\usepackage{etoolbox}
\usepackage[numbers,sort&compress]{natbib} 
\usepackage{setspace}
\usepackage{authblk}

\usepackage{titlesec}

\newcommand{\beginsupplement}{
    \setcounter{section}{0}
    \renewcommand{\thesection}{S\arabic{section}}
    \setcounter{figure}{0}
    \renewcommand{\thefigure}{S\arabic{figure}}
    \setcounter{table}{0}
    \renewcommand{\thetable}{S\arabic{table}}
    \setcounter{equation}{0}
    \renewcommand{\theequation}{S\arabic{equation}}
}

\titleformat{\section}
  {\large\bfseries\uppercase}  
  {\thesection.}{1em}{}

\title{Triangular cross-section grating couplers for integrated quantum nanophotonic hardware in silicon carbide}
\author[1]{Pranta Saha}
\author[1,2]{Alex H. Rubin}
\author[1]{Sridhar Majety}
\author[3]{Scott Dhuey}
\author[1]{Marina Radulaski}
\affil[1]{Electrical and Computer Engineering Department, University of California, Davis, CA 95616, USA}
\affil[2]{Department of Physics and Astronomy, University of California, Davis, CA 95616, USA}
\affil[3]{The Molecular Foundry, Lawrence Berkeley National Laboratory, 1 Cyclotron Rd, Berkeley, CA 94720, USA}
\date{}

\begin{document}
\maketitle

\begin{abstract}
We design, fabricate, and characterize fishbone grating couplers for triangular cross-section photonics in silicon carbide compatible with color center integration. The periodic and aperiodic grating coupler designs are optimized to outcouple up to 31\% of light in the fundamental TE mode of a triangular waveguide. The devices are fabricated using an ion beam etching process in a 4H-SiC sample implanted with NV center ensembles. The room-temperature transmission and the cryogenic NV center photoluminescence collection measurements indicate experimental grating coupler efficiency of up to 24\%. This result provides a scalable method to efficiently extract color center light from SiC quantum nanophotonic devices to free-space optics.
\end{abstract}

\section{\label{sec1}Introduction}
Quantum networks require a medium that can store quantum information for a long time and provide an interface to transfer the information reliably over long distances. Photons are excellent quantum information carriers as they experience low loss and less decoherence when transmitted through optical fiber networks, making it easier to integrate quantum networks into existing infrastructure \cite{kimble2008network1,monroe2002network2,azuma2023network3}. Color centers, luminous defects in wide band gap semiconductors \cite{norman2021novel}, have sparked interest within the quantum community due to their spin-photon interface where the electron or nuclear spin acts as a stationary qubit and spin-entangled photon acts as a flying qubit \cite{hensen2015loophole}. For quantum networks and distributed quantum information processing (QIP), color centers in silicon carbide (SiC) are particularly interesting as they generate spin-entangled photons near telecom band wavelengths \cite{majety2022jap} with seconds-long spin coherence times \cite{anderson2022five} and ultra-narrow inhomogeneous broadening \cite{cilibrizzi2023ultra}. Moreover, compatibility with silicon-based device fabrication processes and availability of high-purity wafer-scale substrates make SiC a potential platform for scalable QIP hardware. 

Color centers are required to be integrated with nanophotonic devices to perform on-chip QIP \cite{castelletto2022silicon,majety2022jap}. In this regard, suspended photonics paves the way for higher optical confinement due to maximum refractive index contrast with the surrounding medium. Previous efforts in fabricating color center integrated SiC suspended devices either limited the performance of color center optical coherence \cite{lukin2020integrated} or encountered scaling issues \cite{majety2022jap}. As color centers exhibit excellent spectral stability in bulk substrates \cite{udvarhelyi2019spectrally}, the angle-etching method has emerged as a promising technique to fabricate bulk freestanding nanostructures for color center integration.  Angle-etching can be implemented either by a Faraday cage or by ion beam etching (IBE). In this method, ions are directed at an angle toward the substrate to undercut the nanophotonic structures which results in a triangular cross-section profile. SiC color center photonics and spintronics in this non-conventional triangular geometry have demonstrated robust performance for QIP applications \cite{majety2021quantum,babin2022fabrication,majety2023snspd,saha2023utilizing,majety2024bs}.
\begin{figure}
    \centering
    \includegraphics[width=0.8\textwidth]{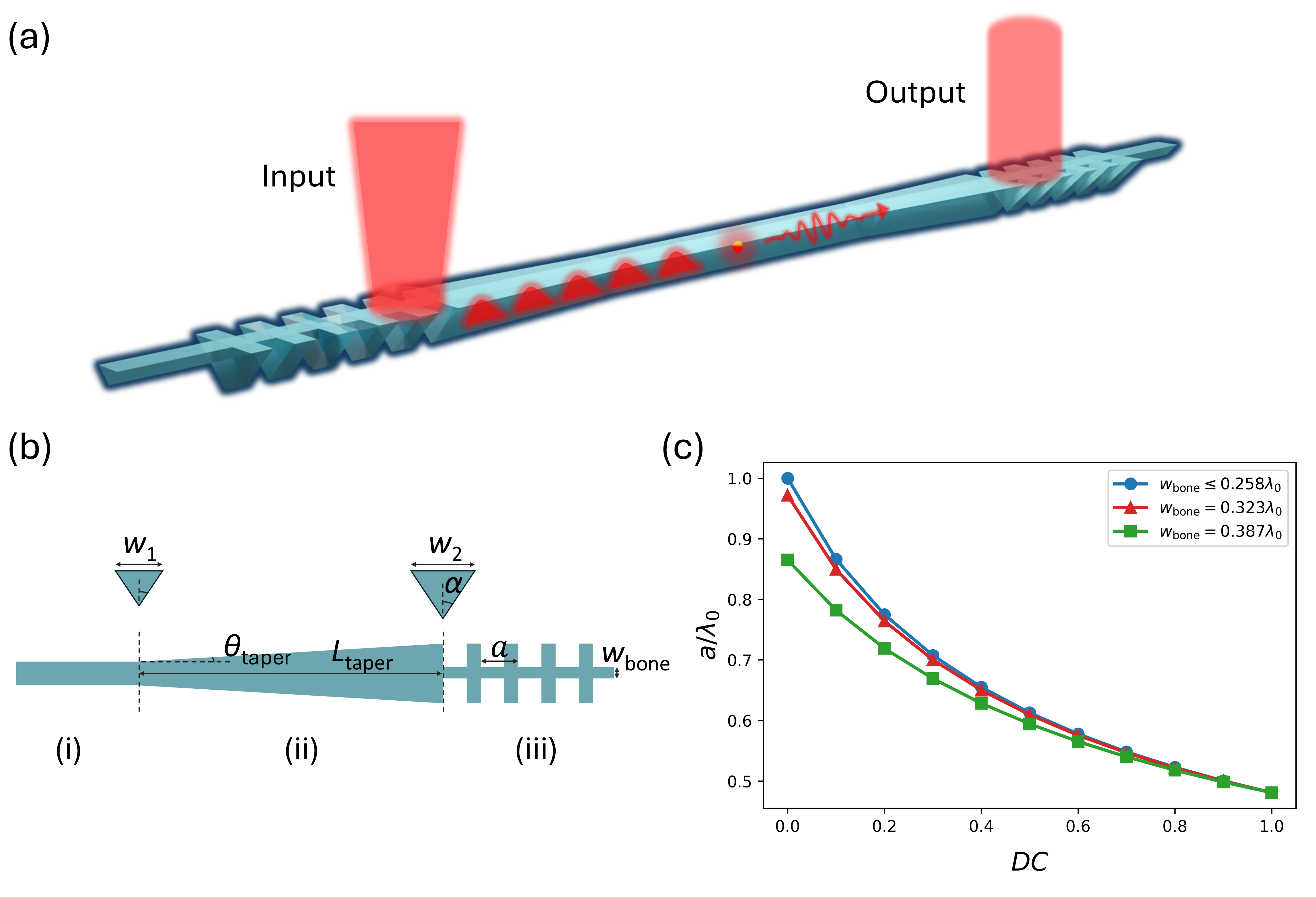}
    \caption{\label{fig1}(a) A perspective figure of the triangular cross-section grating couplers (GCs) used for the characterization of color center coupling with integrated nanophotonic hardware. The red sphere at the center of the waveguide represents the color center. (b) Top view of triangular cross-section fishbone GC with three regions: (i) waveguide, (ii) adiabatic taper, and (iii) fishbone grating. (c) Grating period $a$ and duty cycle $DC$ which satisfy the 1st order vertical grating phase matching condition for $w_\mathrm{bone}$ = $0.258\lambda_\mathrm{0}$, $0.323\lambda_\mathrm{0}$, $0.387\lambda_\mathrm{0}$.}
\end{figure}

Recently, we have developed a reactive ion beam etching (RIBE) process for wafer-scale integration of triangular cross-section photonic devices with color centers in SiC \cite{majety2024wafer}. An on-chip wafer-scale testing mechanism is required for characterizing color center interaction with triangular shaped nanophotonic hardware. Although edge couplers offer highly efficient coupling over a large bandwidth, they do not support wafer-level testing and have complex post-fabrication steps with limited access to the circuits \cite{cheng2020rev1,marchetti2019rev2}. Vertical grating coupler (GC) is a convenient choice for flexible integration with other photonic elements on the same chip, allowing for wafer-scale testing. Fiber-based GCs exhibit excellent coupling efficiency at the cost of large footprints and precise fiber alignment \cite{marchetti2019rev2}. As a cryogenic environment is required for color center characterization \cite{norman2024icecap}, free-space GCs in combination with a microscope objective have proven to be a very practical solution \cite{faraon2008dipole,zhou2018high}. Typical microscopy systems have a higher numerical aperture (NA) compared to optical fibers and provide wider bandwidth \cite{zhu2017ultra}. Moreover, due to the absence of mode mismatch issues between the fiber core and the GC, free-space GCs can be optimized for a compact layout.

In this work, we design, fabricate, and characterize SiC triangular cross-section grating couplers compatible with quantum-grade color center substrates. We first model periodic and aperiodic GC designs using the Finite-Difference Time-Domain (FDTD) method. We then use e-beam lithography, Reactive Ion Etching (RIE) and Reactive Ion Beam Etching (RIBE) processes to undercut devices with triangular cross-section. We characterize devices in room-temperature transmission and cryogenic photoluminescence setups and obtain waveguide propagation loss and GC efficiencies, indicating that fishbone grating couplers are a suitable device to outcouple light from triangular color center quantum photonics to free-space optics in a scalable process.

\section{\label{sec2}Triangular cross-section fishbone grating coupler}
In this section, we analyze the design parameters for triangular cross-section suspended GC in SiC. The grating coupler, illustrated in Fig. \ref{fig1}(a), consists of (i) a triangular waveguide, (ii) an adiabatically tapered region, and (iii) a fishbone-like grating structure. We choose the fishbone grating design in accordance with the fabrication compatibility of the RIBE process to produce wafer-scale triangular cross-section devices. Fig. \ref{fig1}(b) shows the top view schematic of the GC with waveguide width $w_\mathrm{1}$, etch angle (half angle at the apex) $\alpha$, taper angle $\theta_\mathrm{taper}$, taper length $L_\mathrm{taper}$, maximum taper width $w_\mathrm{2}$, grating period $a$, and the central bone width $w_\mathrm{bone}$. The refractive index of SiC is kept $n_\mathrm{SiC} =2.6$ throughout the analysis. We parameterize the scales w.r.t. wavelength of interest $\lambda_\mathrm{0}$ to make the design scalable for the variety of color centers SiC hosts. We optimize each component individually to achieve the most efficient design. 

\subsection{\label{sec2:p1}Single-mode waveguide}
Single-mode propagation is an important requirement in the waveguide design for performing on-chip QIP \cite{caves1994quantum,humphreys2013linear}. For each $\alpha$, there exists an optimal width for single-mode propagation in the triangular cross-section waveguides \cite{babin2022fabrication}. In this work, we choose $\alpha = 45^{\circ}$ as it offers higher confinement \cite{saha2023utilizing} and coupling \cite{babin2022fabrication,majety2024bs} compared to other etch angles that have been fabricated with the RIBE process in SiC \cite{majety2024wafer}. Although the proposed design is a 1D-GC structure in which we can optimize only for a single state of polarization \cite{marchetti2019rev2}, we choose $w_\mathrm{1} = 0.548 \lambda_\mathrm{0}$ to obtain high coupling for both the fundamental transverse electric (f-TE) and the fundamental transverse magnetic (f-TM) modes of the waveguide (See supplementary material S1.1 for details).  

\subsection{\label{sec2:p2}Adiabatic taper}
An adiabatically tapered region is instrumental to coupling light in and out of the chip \cite{son2018taper1}. It allows for a smooth transition of light from the waveguide mode to the Gaussian mode by gradually changing the size and shape of the optical mode along the taper. The slow transition of the effective refractive index ($n_\mathrm{eff}$) of the waveguide cross-section ensures minimal mode conversion to higher-order or radiation modes resulting in high coupling efficiency. Traditional fiber-coupled GCs require very long taper regions to match the mode field diameter of the single-mode optical fibers \cite{cheng2020rev1}. On the other hand, light coupling through the microscope objective enables very compact designs for such tapers as the focused Gaussian beam spot size becomes orders of magnitude smaller compared to the optical fiber mode. As a result, we are able to design a highly efficient compact adiabatic taper with \{$\theta_\mathrm{taper}$, $w_\mathrm{2}$, $L_\mathrm{taper}$\} = \{$2^{\circ}$, $0.774\lambda_\mathrm{0}$, $3.225\lambda_\mathrm{0}$\} (See supplementary material S1.2 for details). 

\subsection{\label{sec2:p3}Fishbone grating}
Refractive index variation with periodic configuration in the direction of light propagation is the most preferred approach for breaking the device symmetry to achieve out-of-plane coupling through diffraction\cite{cheng2020rev1}. Our fishbone grating is constructed by varying the width of the triangular cross-section periodically to implement refractive index contrast. The central bone ensures the mechanical stability of the suspended structure. The diffraction mechanism for a GC is governed by Bragg phase matching condition: $N_\mathrm{eff} - n_\mathrm{air}$sin$\theta$ = $m\lambda_\mathrm{0}/a$, where $N_\mathrm{eff}$ is the effective index of the guided mode in the grating medium, $n_\mathrm{air}$ is the refractive index of air, $\theta$ is the diffraction angle, $m$ is the grating order, $\lambda_\mathrm{0}$ is the target wavelength, and $a$ is the grating period. For 1D-GC, the 1st order diffracted mode is vertically emitted from the grating \cite{marchetti2019rev2}. With $\theta = 0^{\circ}$ and $m=1$, we can achieve vertical outcoupling and the phase matching condition reduces to $a=\lambda_\mathrm{0}/N_\mathrm{eff}$. Due to the 1D periodicity, we can optimize for only one polarization state, in our case the f-TE mode. We estimate the $N_\mathrm{eff}$ using the analytical expression \cite{zhang2015subwavelength}: $N_\mathrm{eff} = \sqrt{(1-DC) n_\mathrm{1}^2 + DC n_\mathrm{2}^2}$, where $DC$ is the duty cycle, or the fill factor, of the wider section, and $n_\mathrm{1}$ and $n_\mathrm{2}$ are the effective indices of the triangular cross-section with widths $w_\mathrm{bone}$ and $w_\mathrm{2}$, respectively. The grating period $a$ and duty cycle $DC$ that satisfy the phase matching condition for different $w_\mathrm{bone}$ values are presented in Fig. \ref{fig1}(c). We observe that there exists no physical mode in the triangular cross-section for $w_\mathrm{bone} \leq 0.258\lambda_\mathrm{0}$ at $\lambda_\mathrm{0}$, which means the effective index of the narrow regions is $n_\mathrm{1} = 1$. Therefore, we choose $w_\mathrm{bone} = 0.258\lambda_\mathrm{0}$ for obtaining maximum index contrast.  
\begin{figure}[H]
    \centering
    \includegraphics[width=0.8\textwidth]{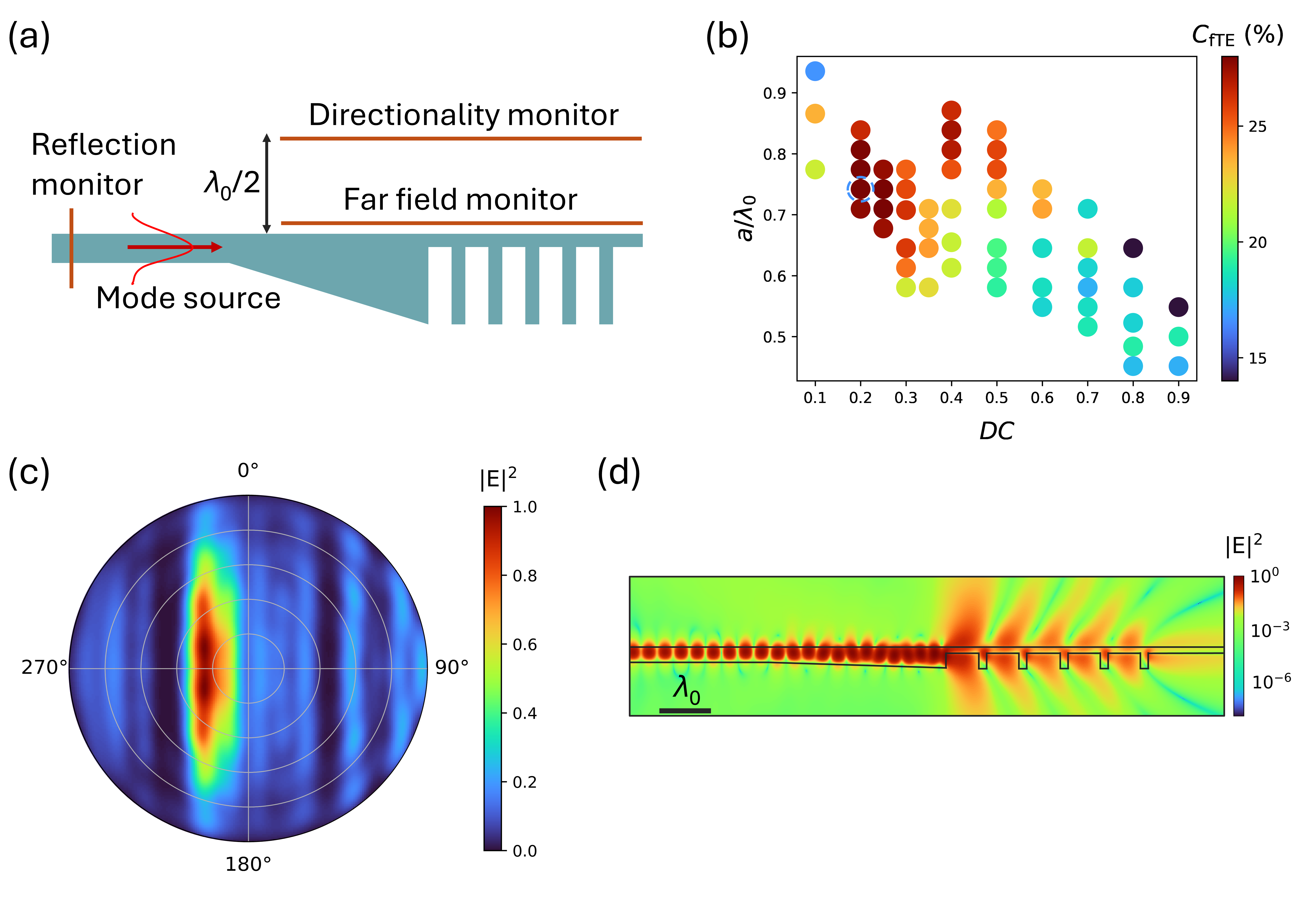}
    \caption{\label{fig2}(a) Side view of the FDTD simulation setup for calculating collection efficiency. (b) Collection efficiency ($C_\mathrm{fTE}$) for the f-TE mode injection from the waveguide with variation in $DC$ and $a$. The dashed (blue) circle indicates the most optimized design for the maximum $C_\mathrm{fTE}$. (c)-(d) The far field radiation pattern and the side view cross-section $|\mathrm{E}|^2$ distribution, respectively, of the triangular GC with $DC = 0.2$ and $a = 0.742\lambda_\mathrm{0}$, marked with blue dashed circle in (b).}
\end{figure}

\section{\label{sec3} Design Optimization}
The GC design optimization is focused on enhancing the collection efficiency from color center integrated nanophotonic devices for characterizing light-matter interaction. Fig. \ref{fig2}(a) shows the side view of the 3D FDTD simulation setup for calculating the collection efficiency. We use a mode source to inject the f-TE mode of the waveguide (width $w_\mathrm{1}$) at $\lambda_\mathrm{0}$ into the grating. A field and power monitor is placed above the triangular cross-section GC at a $\lambda_\mathrm{0}/2$ distance for measuring the fraction of power going in the upward direction. As we are interested in maximizing the light collection through a microscope objective, we add another field and power monitor very close to the top surface of the GC to calculate the far field radiation pattern and subsequently the fraction of power collected by an objective with a given NA (0.65 in this work). The collection efficiency is obtained by multiplying the power fractions calculated from these two monitors. We use a field and power monitor before the source to check the amount of light reflected back into the waveguide.

\begin{figure}[H]
    \centering
    \includegraphics[width=0.8\textwidth]{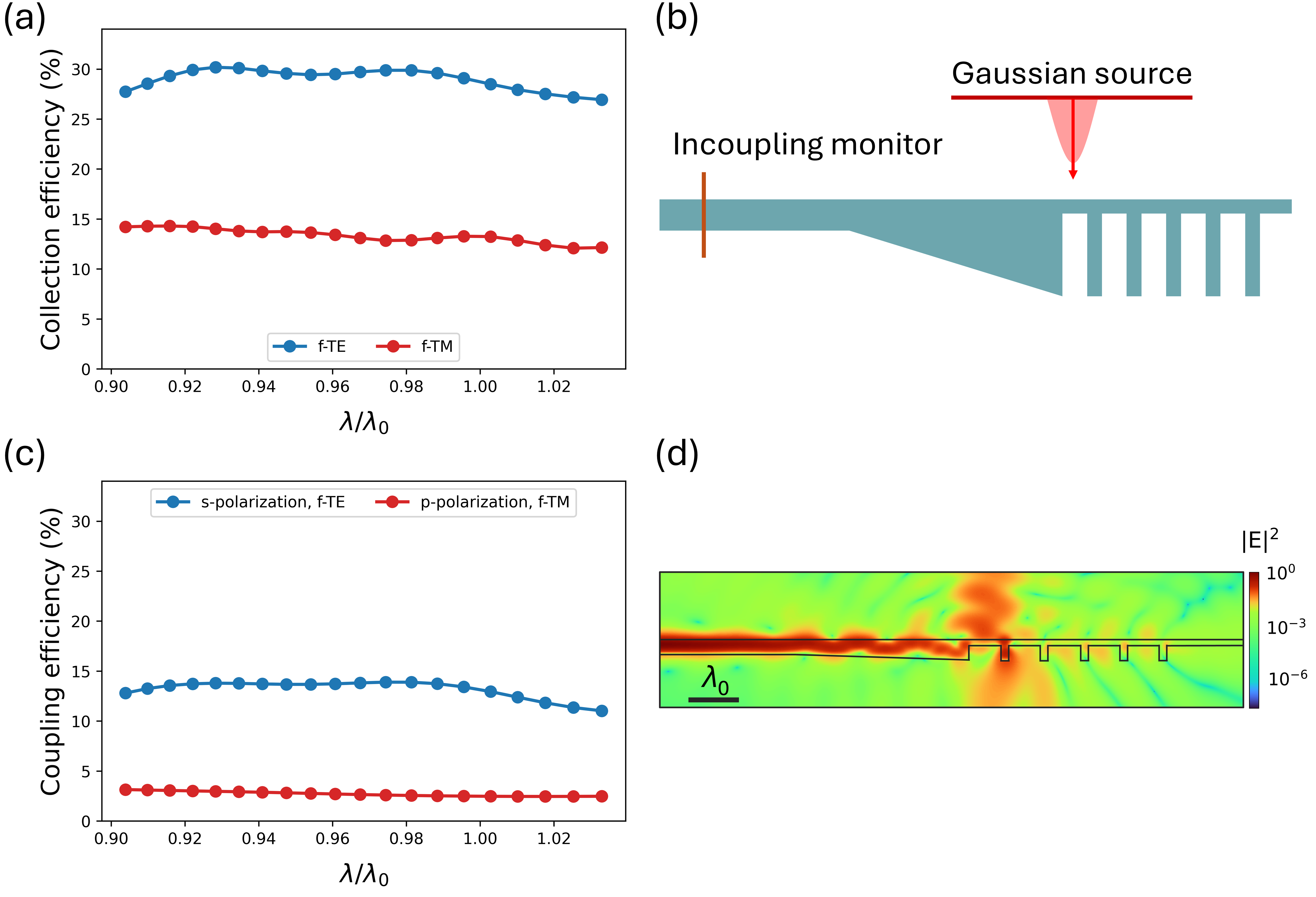}
    \caption{\label{fig3}(a) Collection efficiency of the optimized design with periodic grating for the f-TE and f-TM modes as a function of normalized wavelength $\lambda/\lambda_\mathrm{0}$. (b) Side view of the FDTD simulation setup for calculating the incoupling efficiency. (c) Incoupling efficiency for the s-polarized and p-polarized Gaussian beams as a function of normalized wavelength $\lambda/\lambda_\mathrm{0}$. (d) The side view cross-section $|\mathrm{E}|^2$ distribution of the triangular GC for an s-polarized Gaussian beam focused from above at $\lambda_\mathrm{0}$.}
\end{figure}

For a given design, the collection increases with the number of grating periods and saturates for 5 unit cells. As a result, we use 5 grating periods for all the GCs in our analysis. Fig. \ref{fig2}(b) exhibits the collection efficiencies ($C_\mathrm{fTE}$) from the injected f-TE waveguide mode with variation in $DC$ and $a$. We observe that reflection decreases with the increase in $C_\mathrm{fTE}$. The maximum $C_\mathrm{fTE}$ is $\sim 29\%$, acquired from a design with $DC =0.2$ and $a = 0.742\lambda_\mathrm{0}$. The obtained $C_\mathrm{fTE}$ value is comparable to another compact GC design realized in silicon nitride for microscope objective collection where the theoretical coupling efficiency reaches up to $40\%$ ($60\%$ with a reflective mirror coating on the substrate) \cite{zhu2017ultra}. Notably, the most efficient design does not satisfy the phase matching condition which can stem from the broken symmetry of the triangular profile compared to rectangular structures. This phenomenon is observed in another theoretical work with triangular cross-section GCs in gallium nitride \cite{hadden2022design}. Fig. \ref{fig2}(c)-(d) show the far field radiation pattern and side view cross-section of electric field intensity ($|\mathrm{E}|^2$) distribution, respectively, at $\lambda_\mathrm{0}$ for the most efficient design. We also simulate the collection efficiency of the optimized design as a function of normalized wavelength $\lambda/\lambda_\mathrm{0}$. The collection efficiency for the f-TM mode at $\lambda_\mathrm{0}$ is $\sim13\%$, lower than half of $C_\mathrm{fTE}$, which is expected. Conventional 1D grating couplers have to deal with bandwidth issues \cite{andreani2016bw1,vitali2023bw2} whereas the triangular cross-section GC response appears flat over a wide bandwidth, meaning the same design can be utilized for studying different zero-phonon line (ZPL) emissions from a color center in SiC. To characterize the incoupling efficiency of the optimized design, we employ a Gaussian source focused from above by an objective with 0.65 NA and a power monitor in the waveguide region as shown in Fig. \ref{fig3}(b). For higher incoupling, the source needs to be focused on the first grating period. Focusing on other unit cells results in lower incoupling as a large amount of light readily couples to the guided mode of the adiabatic taper as seen in Fig. \ref{fig3}(d). The s-polarized Gaussian beam mostly couples to the f-TE mode while the p-polarized beam couples mostly to the f-TM mode. Although we observe some initial coupling to high-order modes, they diminish as the incoupled light propagates through the waveguide. Fig. \ref{fig3}(c) illustrates the broadband response for the incoupling efficiency of the optimized design. The incoupling efficiency is much lower than outcoupling as our optimization technique prioritizes the collection efficiency. 

\begin{figure}[H]
    \centering
    \includegraphics[width=\textwidth]{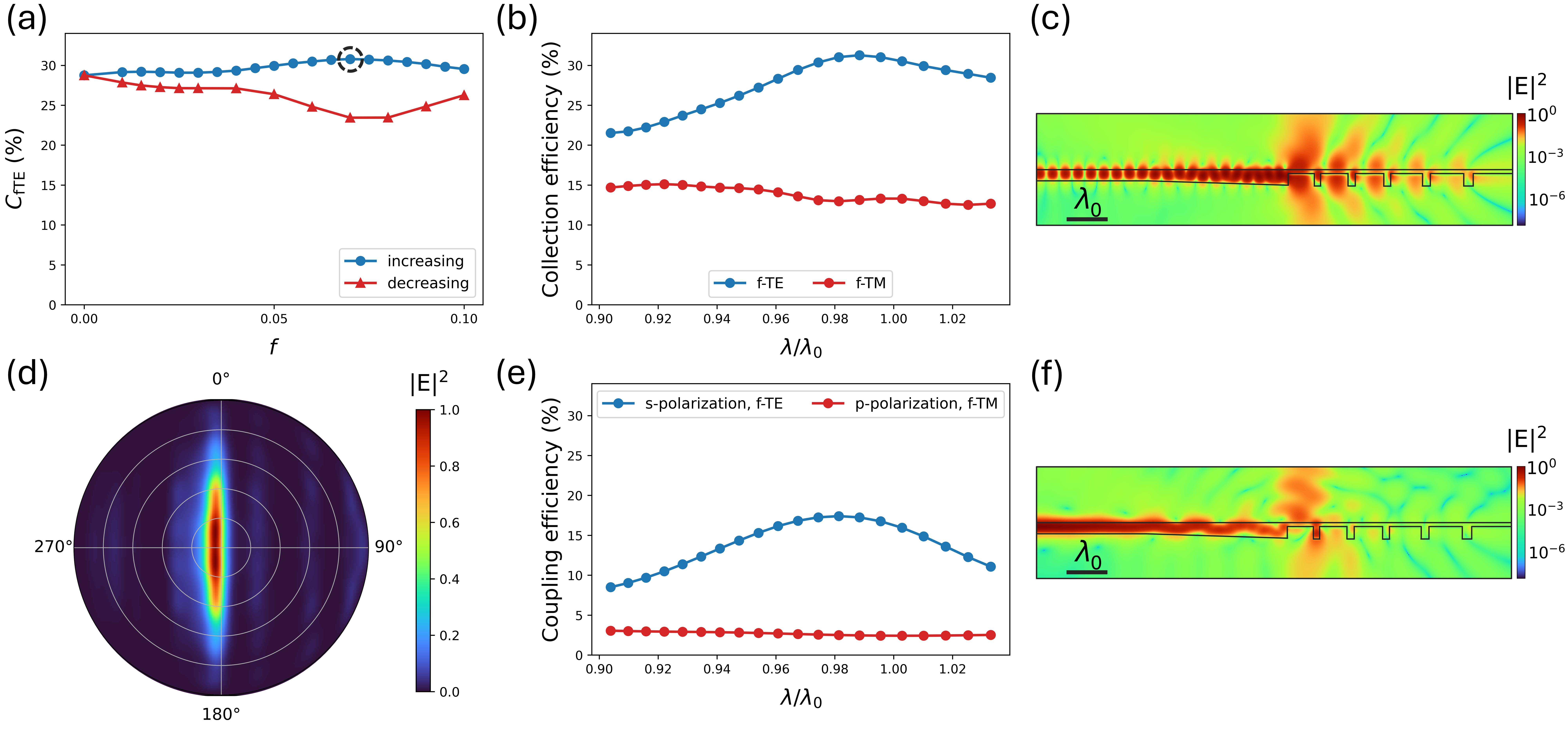}
    \caption{\label{fig4}(a) $C_\mathrm{fTE}$ with variation in aperiodic factor $f$ for linear increase and decrease of $a$. The dashed (black) circle shows the most efficient aperiodic design for the maximum $C_\mathrm{fTE}$. (b) Collection efficiency of the optimized design with aperiodic grating, marked with a black circle in (a), as a function of normalized wavelength $\lambda/\lambda_\mathrm{0}$. (c)-(d)  The side view cross-section $|\mathrm{E}|^2$ distribution and the far field radiation pattern, respectively, of the triangular aperiodic GC with $f = 0.07$ at $\lambda_\mathrm{0}$. (e) Incoupling efficiency for the s-polarized and p-polarized Gaussian beams as a function of normalized wavelength $\lambda/\lambda_\mathrm{0}$ for the optimized aperiodic GC. (f) The side view cross-section $|\mathrm{E}|^2$ distribution of the triangularly shaped aperiodic GC for an s-polarized Gaussian beam focused from above at $\lambda_\mathrm{0}$.}
\end{figure}

Non-uniform GCs have proven to be very effective in enhancing the overall efficiency \cite{marchetti2019rev2,ding2013apo1,lee2016apo2}. Non-uniformity is generally introduced by linearly apodizing the $DC$ or $a$ along the direction of propagation. In search of the most optimized design, we opt for a similar technique. Keeping $a = 0.742\lambda_\mathrm{0}$ constant, we linearly change the $DC$ of each unit cell with the following equation: $DC_i = DC_\mathrm{0} \pm r \times i$, where $r$ is the apodization factor, $DC_\mathrm{0} = 0.2$, and $i = 0, 1, 2, 3, 4$. However, we observe the $C_\mathrm{fTE}$ slowly decreasing for both cases with variation in $r$ from 0 to 0.02. On the other hand, we linearly change $a$ for each grating section, with constant $DC = 0.2$, by the following equation:  $a_i = a_\mathrm{0} (1 \pm f \times i)$, where $f$ is the aperiodic factor, $a_\mathrm{0} = 0.742\lambda_\mathrm{0}$, and $i = 0, 1, 2, 3, 4$. Fig. \ref{fig4}(a) demonstrates that $C_\mathrm{fTE}$ gradually increases (decreases) with increasing (decreasing) $a$ and starts to decrease (increase) after reaching an inversion point. We notice that even for the most optimized aperiodic design, with $f = 0.07$, $C_\mathrm{fTE}$ only increases by $\sim2\%$. However, we observe a significant change in the broadband response, Fig. \ref{fig4}(b), and the far field radiation pattern, Fig. \ref{fig4}(d), of the aperiodic GC. The far field radiation appears less dispersive compared to the periodic design. While collection and incoupling efficiencies for the f-TM mode remain invariant for the aperiodic optimization, the broadband response for the f-TE mode becomes a normal distribution. Based on the specific application requirement, one can adopt either the periodic or the aperiodic design as both designs exhibit comparable collection efficiencies.  

\section{\label{sec4} Experimental Results}
Designed grating couplers can readily be implemented with the wafer-scale nanophotonic fabrication processes. As illustrated in Fig. \ref{fig1}(a), a suspended triangular waveguide has grating couplers on its both ends. One coupler is used to direct the laser into the waveguide and excite the color center with off-resonant excitation. The other one couples both the excitation light and the color center emission out into free-space optics where the excitation beam is filtered out. With this mechanism, light-matter interaction in the triangular geometry can be experimentally characterized for all suspended active and passive photonic devices. 


\subsection{\label{sec4:p1} Nanofabrication Process}
\begin{figure}[H]
    \centering
    \includegraphics[width=\textwidth]{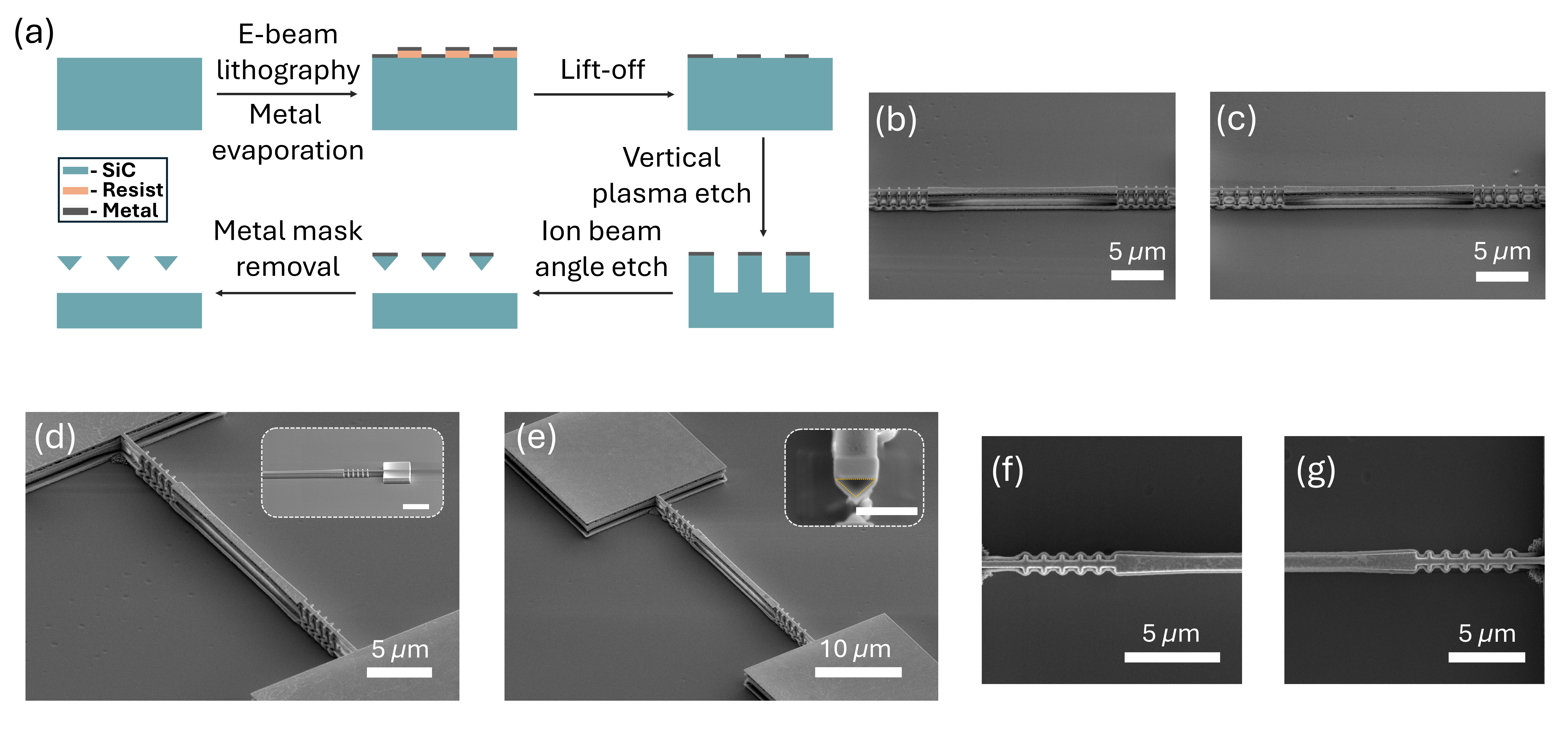}
    \caption{\label{fig5}(a) Nanofabrication process flow of triangular cross-section fishbone GC in SiC. (b)-(c) Side view and (d)-(e) Perspective view SEM images of the triangular cross-section periodic and aperiodic GCs, respectively, with a 52$^\circ$ tilt. (f)-(g) Top view SEM images of triangular cross-section periodic and aperiodic GCs, respectively. Inset (d) shows the side view (52$^\circ$ tilt) of a periodic GC after vertical etch, scale bar indicates 5 $\mu$m. Inset (e) shows the FIB cross-section SEM image (52$^\circ$ tilt) of a triangular waveguide with $\alpha \simeq 42^\circ$, scale bar indicates 1 $\mu$m. All the SEM images, except inset (e), are obtained with the nickel hard mask on top of the devices.}
\end{figure}

We implement the key step in the fabrication of the studied GC designs in SiC using the RIBE process \cite{majety2024wafer}. We choose $\lambda_\mathrm{0} = 1310$ nm, considering the broadband response, for both the periodic and aperiodic GC fabrication. Fig. \ref{fig5}(a) shows the nanofabrication process flow. For generating nitrogen vacancy (NV) centers, before nanofabrication, each 4H-SiC sample is irradiated with $^{14}$N$^{+}$ ions with a dose energy of 65 keV and a fluence of 1 $\times$ 10$^{14}$ cm$^{-2}$ at room temperature \cite{majety2024wafer, wang2020coherent}. The implantation is performed commercially through CuttingEdge Ions, LLC. The dose energy results in a peak nitrogen concentration at $\sim$115 nm depth from the surface, calculated by stopping and range of ions in matter (SRIM) simulations, close to the highest $|\mathrm{E}|^2$ point in the simulated triangular cross-section devices for the best coupling to the waveguide modes. The samples are subsequently annealed at 1050$^\circ$C in a nitrogen atmosphere for one hour to activate the NV centers \cite{majety2024wafer, wang2020coherent}. 

We use electron beam lithography for defining the GC patterns on the NV center implanted 4H-SiC samples. Then we deposit 120 nm thick nickel as a hard mask using electron beam evaporation and transfer the GC patterns on to the mask through a lift-off process. To obtain a deeper undercut ($\sim$1.5 $\mu$m etch depth), we perform 3 minutes of inductively coupled plasma-reactive ion etching (ICP-RIE) with  SF$_\mathrm{6}$ and O$_2$ chemistry, which has higher etch rate and selectivity than the RIBE process \cite{norman2024icecap}. The triangular cross-section waveguides and GCs are then fabricated by the RIBE process using SF$_\mathrm{6}$, O$_2$, and Ar gas ions, with the substrate tilt angle set to $60^{\circ}$ to achieve $\alpha = 45^{\circ}$ \cite{majety2024wafer} and overall device undercut $>$1.6 $\mu$m. However, after the nickel hard mask removal, we notice that most of the devices break near the attachment point between the thin ($\sim$300 nm) triangular cross-section bone and the square pads. Previous fabrication efforts in triangular diamond nanobeams faced similar difficulties after mask removal \cite{bayn2014fabrication}. On a subsequent attempt, we handle the sample with extreme caution during mask removal which results in a higher fabrication yield ($\sim$70\%).

\subsection{\label{sec4:p2} Transmission Measurements}
\begin{figure}[H]
    \centering
    \includegraphics[width=0.8\textwidth]{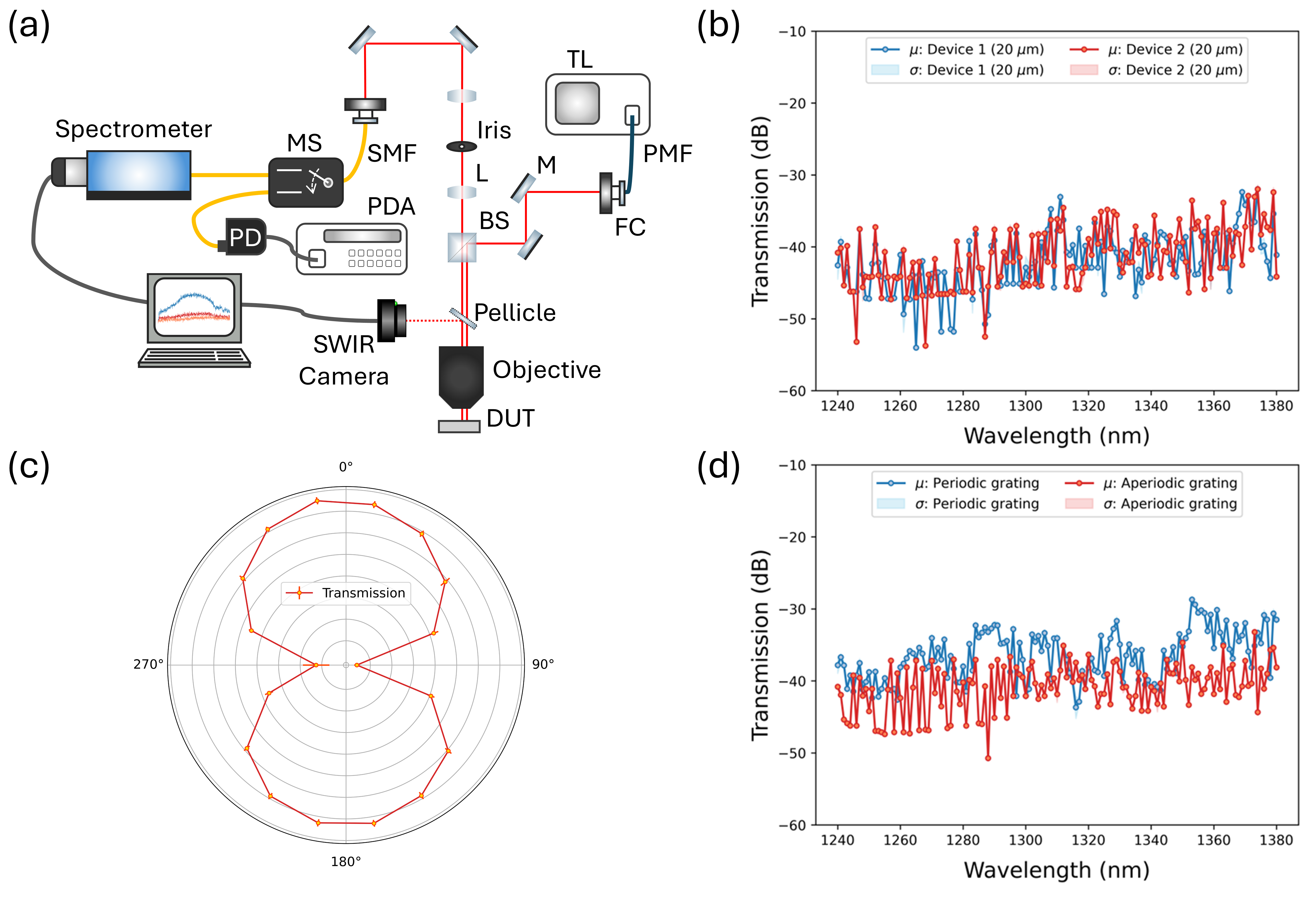}
    \caption{\label{fig6}(a) A sketch of the experimental setup used to characterize fabricated devices; TL: tunable laser, PMF: polarization-maintaining fiber, FC: fiber coupler, M: mirror, BS: 10:90 (R:T) beamsplitter, L: achromatic doublet lens, SMF: single-mode fiber, MS: manual switch, PD: photo detector, PDA: photodiode amplifier, DUT: device under test, SWIR: short-wave infrared. (b) The spectral profile of transmission efficiency of two identically fabricated devices (periodic GCs connected by a 20 $\mu$m waveguide) showing uniform performance. (c) Transmission efficiency as a function of the polarization of the incoupled light, where 0$^\circ$ corresponds to the polarization orthogonal to the waveguide propagation direction. (d) The spectral profile of transmission efficiency of devices with periodic and aperiodic GCs (waveguide length 10 $\mu$m).}
\end{figure}

To characterize the overall efficiency of the triangular waveguides and fishbone GCs, we use a confocal correlator setup, shown in Fig. \ref{fig6}(a), to spatially filter out the reflection of the input beam at the Fourier plane. A linearly polarized tunable laser (Santec TSL-570-A, 1240-1380 nm) is used to perform the transmission measurements at room temperature. The input beam is mostly s-polarized with a $10^\circ$ tilt (CCW)  due to the slow axis position of the polarization-maintaining fiber (PMF) at the fiber coupler (FC) end. The NA of the microscope objective in the setup is 0.65, similar to the simulations. We employ a short-wave infrared (SWIR) camera to monitor the input and the output beams. The collected beam power is measured by an InGaAs photodetector (PD), Thorlabs DET01CFC, which is connected to a benchtop photodiode amplifier (PDA), Thorlabs PDA200C, to ensure stable output. 

\begin{figure}[H]
    \centering
    \includegraphics[width=\textwidth]{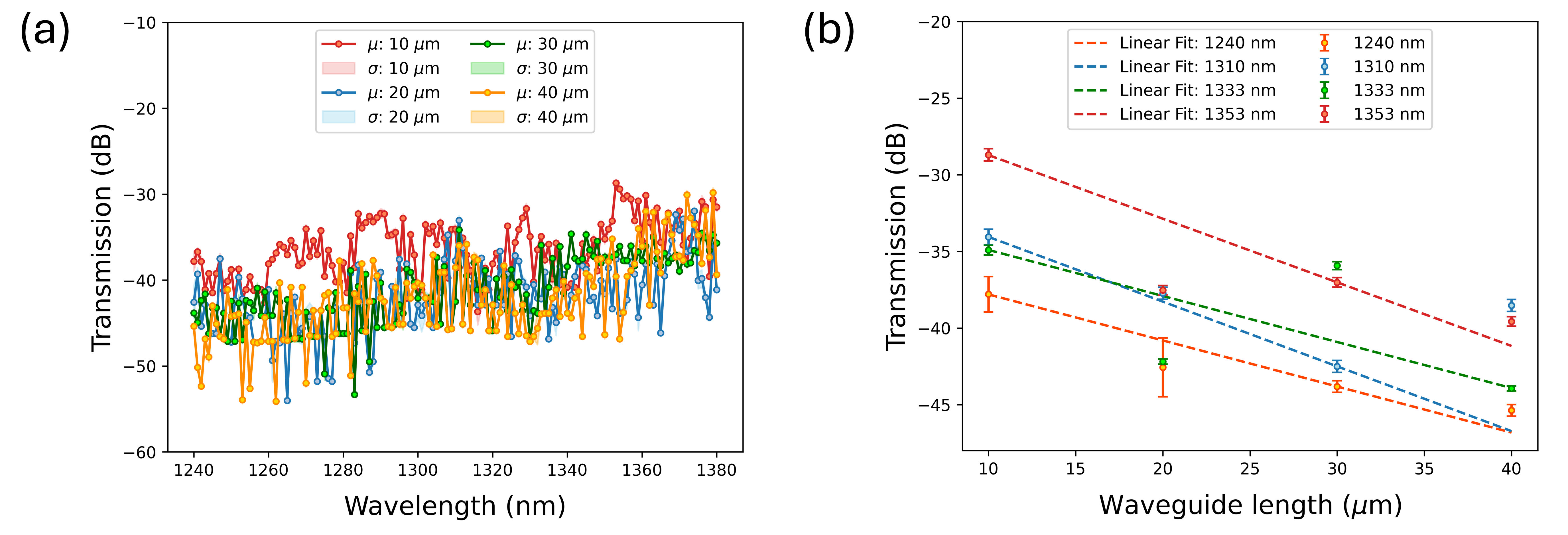}{\centering}
    \caption{\label{fig7}(a) The spectral profile of transmission efficiency through a pair of periodic GCs connected by a waveguide with varying length. (b) Transmission efficiency as a function of the waveguide length for a set of wavelengths indicating waveguide loss in the range [0.3, 0.4] dB/$\mu$m. Error bars indicate the uncertainty in dB.}
\end{figure}

To account for all potential losses arising from both optical and electronic components, we determine the reference input power by measuring the reflected signal from a p-type single-side polished 620 $\mu$m thick test grade silicon substrate. We calculate the input power by normalizing the reflected signal with known reflectance ($R$) of the silicon substrate where $R = \frac{(n-1)^2+k^2}{(n+1)^2+k^2}$ and ($n,k$) values in the 1240-1380 nm range are obtained from literature \cite{schinke2015uncertainty}. To minimize measurement bias, collected power at each wavelength is recorded three times for the silicon reflection and the GC output signals. The mean ($\mu$) and the standard deviation ($\sigma$) of the measurements are calculated by the following equations: $\mu = \frac{1}{N}\sum_{i=1}^{N} x_i$, $\sigma = \sqrt{\frac{1}{N-1}\sum_{i=1}^{N} (x_i - \mu)^2}$ where $x_i$ represents individual measurements and $N$ = 3. We normalize the GC output power by the input power for obtaining the transmission efficiency with $\mu_\mathrm{norm} = \frac{\mu_\mathrm{GC}}{\mu_\mathrm{Si}}$, however, to accurately propagate the measurement uncertainty, we calculate the normalized $\sigma$ with $\sigma_\mathrm{norm} = \mu_\mathrm{norm} \sqrt{\left( \frac{\sigma_\mathrm{GC}}{\mu_\mathrm{GC}} \right)^2 + \left( \frac{\sigma_\mathrm{Si}}{\mu_\mathrm{Si}} \right)^2}$. Although conversion to decibels (dB) is straightforward for $\mu_\mathrm{norm}$, converting propagated uncertainty $\sigma_\mathrm{norm}$ to dB requires careful consideration of error propagation in logarithmic scales \cite{taraldsen2015uncertainty}. The conversion equations used for the measured transmission (dB) presented in Fig. \ref{fig6}, \ref{fig7} are $\mu_\mathrm{dB} = 10\log_{10}(\mu_\mathrm{norm})$ and $\sigma_\mathrm{dB} \approx \frac{10\cdot\sigma_\mathrm{norm}}{\mu_\mathrm{norm}\cdot \ln(10)}$ where we consider $\frac{\sigma_\mathrm{norm}}{\mu_\mathrm{norm}} \ll 1$. 

Transmission efficiency measurements from two identical devices, where each device consists of two identical periodic GCs connected by a 20 $\mu$m triangular waveguide, are presented in Fig. \ref{fig6}(b). The comparable performance of the identical devices highlights the uniformity and reproducibility of the fabrication process. The spectral response exhibits the broadband nature of the fishbone GCs, in good agreement with the simulation results. To verify the polarization response, we insert a half-wave plate in the input beam path and observe a higher transmission for the s-polarized beam compared to the p-polarized beam, shown in Fig. \ref{fig6}(c), consistent with the design optimization for the f-TE mode. In the polar plot, we account for the 10$^{\circ}$ tilt correction of the PMF slow-axis position. Although the aperiodic GCs are expected to outperform the periodic GCs, we observe a slight decrease in the overall transmission, illustrated in Fig. \ref{fig6}(d). On the other hand, the overall transmission efficiencies for both periodic and aperiodic GC devices are orders of magnitude lower compared to the simulation results. Therefore, to evaluate the loss contributions from individual components of the devices, we measure the transmission efficiency of the periodic GC devices as a function of waveguide length, shown in Fig. \ref{fig7}. We choose four wavelengths, covering the operational bandwidth of the devices, and quantify the loss contributions from the individual components, presented in Table \ref{tab:table1}.

\begin{table}[H]
\caption{\label{tab:table1} Efficiency of individual components for the four nominal wavelengths}
\centering
\resizebox{\textwidth}{!}{%
\begin{tabular}{|c|c|c|c|c|c|c|}
\hline
Wavelength&Waveguide&Periodic GCs&Incoupling$^{a}$ (simulation)&Outcoupling$^{a}$ (simulation) & Incoupling$^{b}$ & Outcoupling$^{b}$\\
(nm)&(dB/$\mu$m)&(dB)&(dB)&(dB)&(dB)&(dB)\\
\hline
1240 & -0.30 ± 0.01 & -34.8 ± 0.2 & -8.64 & -5.29 & -21.58 & -13.22\\
1310 & -0.42 ± 0.01 & -29.8 ± 0.1 & -8.88 & -5.45 & -18.47 & -11.35\\
1333 & -0.30 ± 0.02 & -31.9 ± 0.6 & -9.27 & -5.60 & -19.88 & -12.02\\
1353 & -0.41 ± 0.02 & -24.6 ± 0.3 & -9.57 & -5.70 & -15.4 & -9.16 \\
\hline
\end{tabular}%
}
\vspace{1ex}
\begin{flushleft}
\footnotesize
$^{a}$ Incoupling and outcoupling efficiencies are considered only for the f-TE mode, as Fig.~\ref{fig6}(c) shows very low coupling for the f-TM mode compared to the f-TE mode.\\
$^{b}$ Incoupling and outcoupling efficiencies are calculated by applying the method of proportionality between the measured and simulated efficiencies. Uncertainties are neglected as they are orders of magnitude lower than the mean values.
\end{flushleft}
\end{table}

We observe substantial losses [0.3,0.4] dB/$\mu$m in the waveguides, contrary to both the simulation results and previous reports on fabricated triangular nanobeams in diamond \cite{burek2014high, bayn2014fabrication}. Although scattering losses may arise from the surface roughness induced by the lift-off of the metal mask, the losses are too high even for an imperfect waveguide supporting only the fundamental mode(s) \cite{melati2014real}. Such high losses are more consistent with the damage introduced during the ion-implantation process \cite{foster2006optical,shakespeare2021effects}. Considering the implantation energy and fluence, we expect a high density of defects spanning most of the f-TE mode region of the triangular waveguide (See supplementary material S2 for details). Such dense defects can result in high losses due to either Rayleigh scattering, optical absorption, or coupling to the radiation modes. However, due to the long operational wavelengths, Rayleigh scattering losses are expected to contribute minimally. On the other hand, ion implantation is known to alter the real and the imaginary parts of the complex refractive index \cite{baydin2016depth}, while annealing above 1600$^{\circ}$C is not able to fully recover the crystal damage in 4H-SiC introduced by nitrogen ion implantation conditions comparable to this study \cite{gimbert1999nitrogen,piskorski2020investigation}. Since ion implantation facilitates mode conversion into higher-order modes \cite{xu2024mode}, we attribute the losses primarily to radiation, given that the designed waveguides support only the fundamental modes. Consequently, we speculate the incoupling (outcoupling) efficiencies, presented in Table \ref{tab:table1}, are overestimated (underestimated) as incoupling demands more stringent phase matching conditions to the guided modes compared to outcoupling \cite{spuesens2016grating}, which can be the potential reason behind the overall inferior performance of the aperiodic GC devices compared to the periodic ones. Therefore, we proceed with cryogenic photoluminescence (PL) measurements for obtaining a more comprehensive insight.    

\subsection{\label{sec4:p3} Cryogenic Photoluminescence Spectroscopy Measurements}
\begin{figure}[H]
    \centering
    \includegraphics[width=\textwidth]{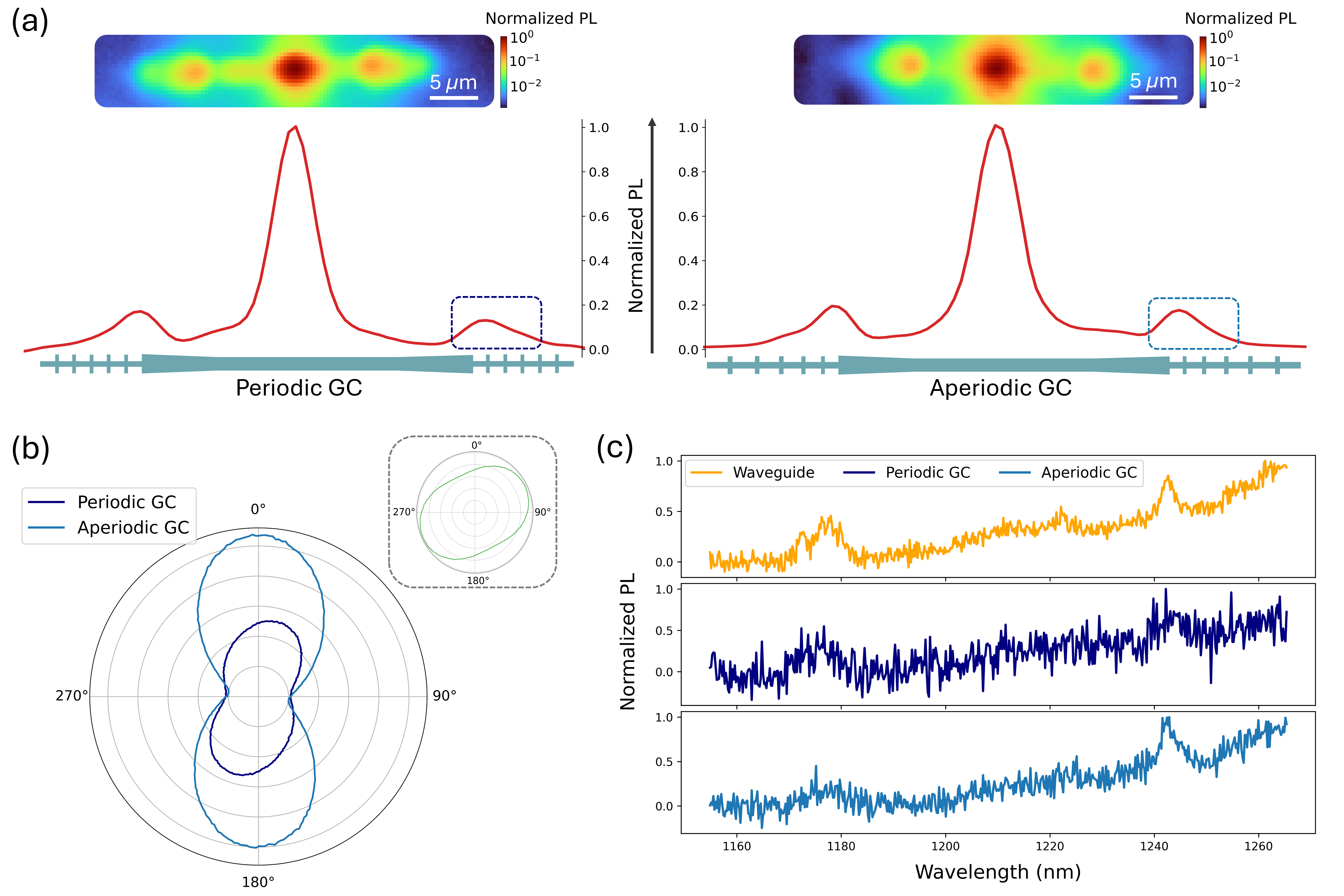}{\centering}
    \caption{\label{fig8}(a) The 2D and 1D scans of normalized photoluminescence (PL) from SiC triangular waveguides terminated by periodic and aperiodic GCs, respectively, with the excitation beam fixed at the center of a 10 $\mu$m waveguide exciting an ensemble of NV centers. The dashed boxes in the 1D PL plots highlight the regions corresponding to the results presented in (b) and (c). (b) PL counts from periodic and aperiodic GCs collected through an analyzer, 0$^\circ$ corresponding to the polarization orthogonal to the waveguide propagation direction. The inset shows PL counts from the NV centers in bulk 4H-SiC, the elliptical shape originating from the non-uniform polarization mixing in the SMF coil of the excitation beam. (c) Normalized PL spectra collected from the center of the waveguide (where NV centers are excited), periodic GC, and aperiodic GC, respectively.}
\end{figure}

We employ the ICECAP 3-in-1 system \cite{norman2024icecap} to conduct the PL characterization of the fabricated devices. All the measurements presented in Fig. \ref{fig8} are performed below 1.6 K, using an above-bandgap CW excitation (785 nm) for the NV centers in 4H-SiC. PL is collected through a higher NA (0.85) objective and a 1150 nm long-pass filter, and subsequently measured using either a superconducting nanowire single-photon detector (SNSPD) or a spectrometer. Fig. \ref{fig8}(a) shows the 2D and 1D PL scans of the fabricated devices where the excitation beam is fixed at the center of the triangular waveguide and a 1-second integration time per point. In the 2D scans, the central red spot corresponds to PL collected directly from the top of the waveguide, whereas the two peripheral bright spots indicate PL extraction via the GCs. The aperiodic GCs yield tighter collection spots relative to the periodic ones, matching the expected far field angular distribution from the simulations. 

Although the GCs connected by the triangular waveguide are symmetric, the 1D PL scans exhibit a slightly asymmetric collection profile. The consistent observation of this phenomenon across both the periodic and aperiodic designs suggests it originates from the sample tilt, likely due to an uneven adhesive application during the cryogenic mounting. Assuming the following conditions: (i) PL (dipole) emission couples equally into the parallel and perpendicular directions w.r.t. the waveguide propagation, and (ii) a 1:1 ratio of axial to basal defects in the NV ensemble, we extract a maximum collection efficiency of $24\%$ (23$\%$) for the aperiodic (periodic) design from the 1D PL scans (See supplementary material S2 and S3 for details). The aperiodic design demonstrates 1-3\% higher efficiency than the periodic, aligning well with the simulated predictions. The polarization characteristics of the collected PL from the periodic and aperiodic GCs, shown in Fig. \ref{fig8}(b), reveal similar f-TM outcoupling for both but a higher f-TE outcoupling for the aperiodic design, as expected from the simulations. Enhanced performance of the aperiodic design is evident from the PL spectra, illustrated in Fig. \ref{fig8}(c). Although all the devices are optimized for $\lambda_0$ = 1310 nm, the PL spectra reflect the inherent broadband nature of this approach.

\section{\label{sec5} Outlook}
In this work, we have expanded the utility of the wafer-scale angle-etching fabrication process to integrate efficient grating couplers with triangular cross-section color center photonics. The process preserves color center optical properties, enabling the implementation of on-chip QIP with integrated quantum photonic mesh architecture. With the addition of the proposed GCs on the same chip, wafer-level testing can be executed on the fabricated devices. Our presented results show that the triangular cross-section fishbone grating design can yield broadband high collection efficiency (theoretical: $\sim31\%$, experimental: $24\%$) for studying the color center coupling with the integrated nanophotonic devices. The uniformity in device performance, along with the consistency with the FDTD simulations, underscores the reliability and reproducibility of the optimized designs using the RIBE process. However, integrating NV centers into SiC triangular photonic devices without compromising the crystal quality remains a nontrivial challenge. Strategies such as hot ion implantation \cite{ngandeu2024hot}, specialized thermal annealing protocol \cite{green2022diamond}, and the use of implantation masks \cite{spinicelli2011engineered} offer promising pathways for minimizing lattice damage and enhancing the emitter performance. In summary, the demonstrated grating couplers can be integrated with mesh photonics, cavity QED systems, and any other process where free-space modes are coupled to quantum photonic devices.

\section*{Supplementary Material}
The supplementary material provides comprehensive explanations of the waveguide and adiabatic taper design optimizations. NV ensemble and other vacancy-type defect integration in the triangular cross-section waveguide is discussed, and PL emission coupling from NV ensemble into the f-TE mode is quantified using FDTD simulations. Combining both simulation and experimental results, the estimated GC efficiency calculations are shown in detail.

\section*{Acknowledgments}
 We acknowledge support from NSF CAREER (Award 2047564) and AFOSR Young Investigator Program (Award FA9550-23-1-0266). Work at the Molecular Foundry was supported by the Office of Science, Office of Basic Energy Sciences, of the U.S. Department of Energy under Contract No. DE-AC02-05CH11231. Part of this study was carried out at the UC Davis Center for NanoMicro Manufacturing (CNM2).  The authors would like to acknowledge the valuable support provided by Elnaz Hamdarsi and CNM2 staff member Vishal Narang.

\section*{Author declaration}
\subsection*{Conflict of Interest}
The authors have no conflict of interest.
\subsection*{Author Contributions}
PS modeled, analyzed, fabricated and characterized devices, AHR performed cryogenic measurements, SM developed the angle etching process, SD performed electron beam lithography, MR supervised the project.

\section*{Data Availability Statement}
The data that support the findings of this study are available upon reasonable request from the corresponding author Pranta Saha (prsaha@ucdavis.edu).

\clearpage
\beginsupplement

\begin{center}
\Large\bfseries SUPPLEMENTARY MATERIAL
\end{center}
\vspace{1em}

\section{DESIGN PARAMETERS}

\subsection{Single-mode waveguide}
\label{S1:p1}
\begin{figure}[ht]
    \centering
    \includegraphics[width=\textwidth]{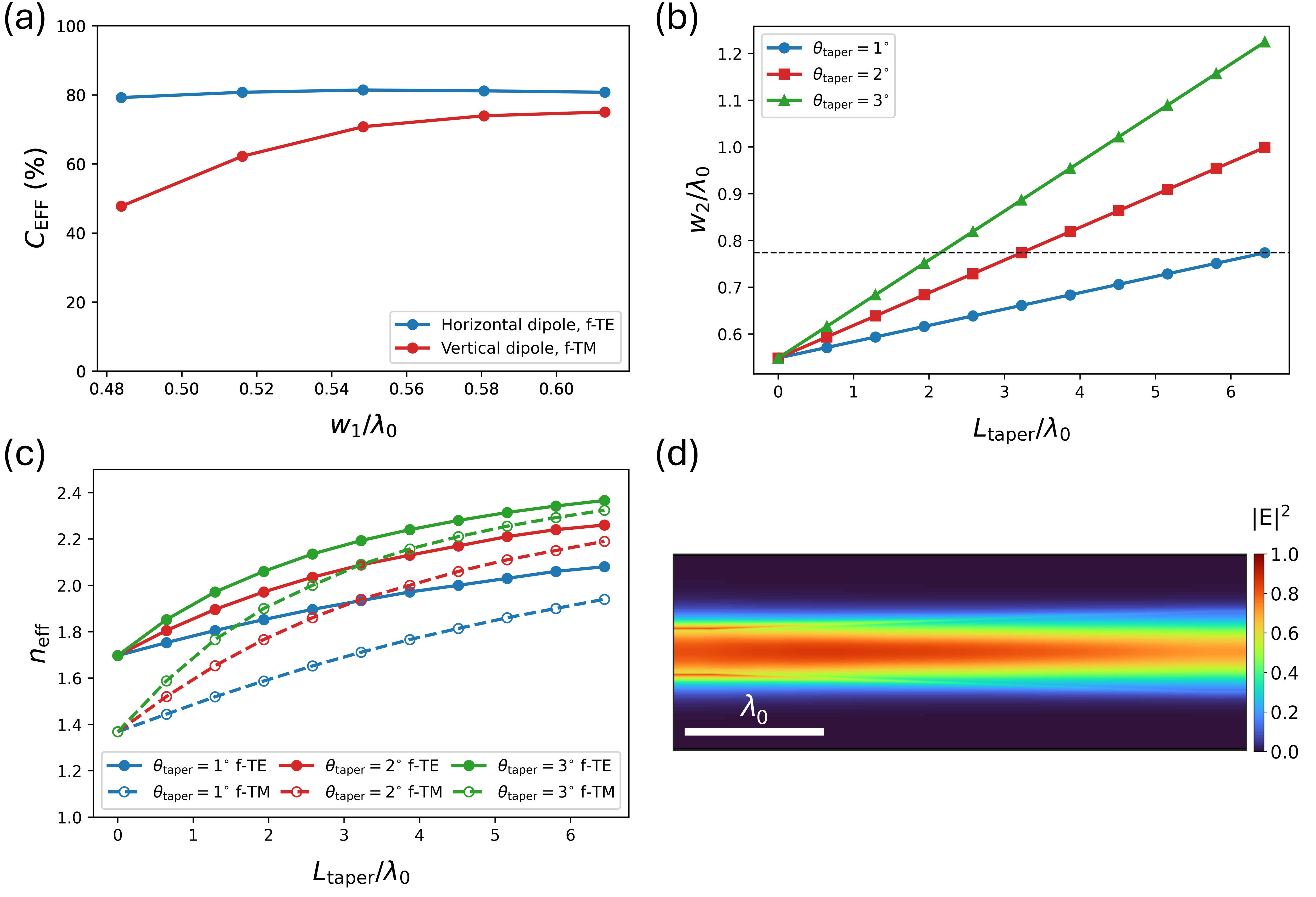}
    \caption{\label{figS1} (a) Coupling efficiency ($C_\mathrm{EFF}$) of horizontal and vertical dipoles into the fundamental transverse electric (f-TE) and the fundamental transverse magnetic (f-TM) modes, respectively, of triangular waveguides with etch angle, $\alpha = 45^{\circ}$.(b) Taper width $w_\mathrm{2}$ w.r.t taper length $L_\mathrm{taper}$ for taper angles $\theta_\mathrm{taper} = 1^{\circ}, 2^{\circ}, 3^{\circ}$. The dashed (black) line shows the fabrication constraint imposed on $w_\mathrm{2}$. (c) Effective refractive indices $n_\mathrm{eff}$ of the f-TE and the f-TM modes along $L_\mathrm{taper}$ for $\theta_\mathrm{taper} = 1^{\circ}, 2^{\circ}, 3^{\circ}$. (d) Top view cross-section of the electric field (E-field) intensity ($|\mathrm{E}|^2$) profile for the adiabatic taper design with $\theta_\mathrm{taper} = 2^{\circ}$, $L_\mathrm{taper} = 3.225\lambda_\mathrm{0}$, and $w_\mathrm{2} = 0.774\lambda_\mathrm{0}$ where the f-TE mode of the waveguide with $w_\mathrm{1} = 0.548\lambda_\mathrm{0}$ is injected into the taper region. The cross-section is taken at the centroid plane of the maximum taper width $w_\mathbf{2}$.}
\end{figure}

Based on the dipole orientation in the SiC crystal lattice, color center emission can couple to fundamental transverse electric (f-TE), fundamental transverse magnetic (f-TM), and other higher-order modes supported by the triangular waveguide. Therefore, we optimize the triangular waveguide width $w_\mathrm{1}$ for maintaining single-mode propagation of both horizontal and vertical electric dipole radiations with high coupling efficiency ($C_\mathrm{EFF}$) using Lumerical FDTD package as shown in Figure \ref{figS1}(a). We position the dipole sources with emission wavelength $\lambda_\mathrm{0}$ at the centroid of the triangle which is the optimum color center position in triangular geometry for the best coupling to the waveguide modes\cite{majety2021quantum,babin2022fabrication,majety2023snspd}. We pre-select the $w_\mathrm{1}$ range using Lumerical MODE (Finite-Difference Eigensolver) package so that the horizontal (vertical) dipole couples only to the f-TE (f-TM) mode.

\subsection{Adiabatic taper}
\label{S1:p2}
In the SOI platform, it has been observed that linear tapers with small taper angle ($\theta_\mathrm{taper}$) result in higher transmission \cite{fu2014taper2}. We apply a similar strategy in designing the triangular cross-section taper region by linearly increasing the taper width for $\theta_\mathrm{taper} = 1^{\circ}, 2^{\circ}, 3^{\circ}$, shown in Figure \ref{figS1}(a). To ensure gradual change in $n_\mathrm{eff}$ of the waveguide mode, we calculate the $n_\mathrm{eff}$ of the f-TE and f-TM modes (mode \#1 and mode \#2, respectively) of the triangular waveguide along $L_\mathrm{taper}$. Figure \ref{figS1}(c) demonstrates that the adiabatic condition for triangular waveguides \cite{burek2017fiber}: $dn/dx < (2\pi/\lambda_\mathrm{0})|n_\mathrm{eff,1}(x) - n_\mathrm{eff,2}(x)|^2$, where $n$ is the effective index of the waveguide cross-section with maximum taper width $w_\mathrm{2}$ and $n_{\mathrm{eff},i}$ is the effective index of the $i$th waveguide mode along $L_\mathrm{taper}$, is satisfied for all three $\theta_\mathrm{taper}$. However, owing to the etch selectivity of the mask in the developed RIBE wafer-scale process\cite{majety2024wafer}, we restrict $w_\mathrm{2}$ to be $0.774\lambda_\mathrm{0}$, $\sim40\%$ larger than $w_\mathrm{1}$, for undercutting both $w_\mathrm{1}$ and $w_\mathrm{2}$ cross-section profiles in a single etch. Although all three $\theta_\mathrm{taper}$ designs exhibit unity transmission for the f-TE mode injection from the waveguide, we choose $\theta_\mathrm{taper} = 2^{\circ}$ leading to $L_\mathrm{taper} = 3.225\lambda_\mathrm{0}$ for $w_\mathrm{2}=0.774\lambda_\mathrm{0}$.

\clearpage
\section{NV CENTER INTEGRATED TRIANGULAR SILICON CARBIDE DEVICES}
\begin{figure}[htbp]
    \centering
    \includegraphics[width=\textwidth]{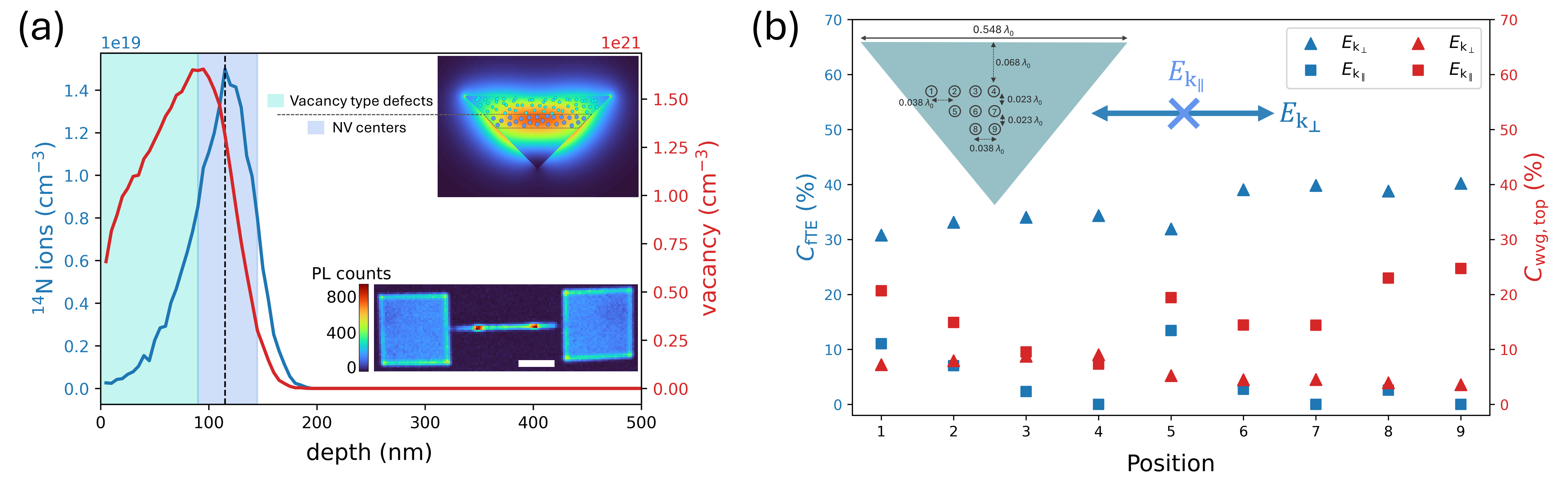}
    \caption{\label{figS2} (a) Depth profile of implanted $^{14}$N$^{+}$ ion and induced vacancy concentrations in 4H-SiC, as simulated by stopping and range of ions in matter (SRIM). The top right inset illustrates the spatial overlap between the f-TE mode and regions containing vacancy-type defects and NV ensemble in the triangular waveguide. The bottom right inset shows the 2D Photoluminescence scan of the NV centers, scale bar indicates 10 $\mu$m. (b) Simulated coupling fraction into the f-TE mode in one direction ($C_\mathrm{fTE}$) and collection efficiency through an objective with 0.85 NA from the top of a triangular waveguide ($C_\mathrm{wvg, top}$) as a function of emitter positions spanning the spatial range of NV ensemble marked in (a).}
\end{figure}

The high implantation energy (65 keV) and fluence (1 $\times$ 10$^{14}$ cm$^{-2}$) result in a high concentration of vacancies ($\mathrm{V_{Si}}$, VV) in the triangular waveguide, as seen in Figure \ref{figS2}(a). Despite annealing for an hour at 1050$^{\circ}$C, the 4H-SiC lattice exhibits only partial recovery \cite{piskorski2020investigation}, which is reflected in the transmission measurements (main text). Annealing in a nitrogen atmosphere suppresses the diffusion of the implanted N-ions, preserving the depth distribution predicted by the stopping and range of ions in matter (SRIM) simulations. However, at 1050$^{\circ}$C, the $\mathrm{V_{Si}}$ and VV defects diffuse towards the surface \cite{mitani2009diffusion}. Based on this analysis, the estimated depth regions corresponding to both vacancy-type defects and NV centers are indicated in Figure \ref{figS2}(a).

We perform Lumerical FDTD simulations to estimate the photoluminescence (PL) emission coupling to one side of the fundamental mode waveguide ($C_\mathrm{fTE}$), as well as the collection from the waveguide top through a 0.85 NA objective ($C_\mathrm{wvg, top}$), under the same conditions as the main text. The dipole moment of axial defects lies in the device plane, while that of basal defects is oriented in a plane tilted by $19^{\circ}$ from the c-axis \cite{norman2024icecap}. To emulate the spatial distribution of NV centers in the triangular waveguide, electrical dipoles are configured parallel ($k_\mathrm{\parallel}$) and perpendicular ($k_\mathrm{\perp}$) to the waveguide propagation direction ($k$) at nine different positions, delineated in Figure \ref{figS2}(b). In the following calculations, we consider a 1:1 ratio of axial to basal defects in the NV ensemble and dipole emission coupling equally into $k_\mathrm{\parallel}$ and $k_\mathrm{\perp}$. From the simulations, we obtain the average values of $C_\mathrm{wvg, top}$ and $C_\mathrm{fTE}$ to be 11.29\% and 20.08\%, respectively. Due to the dipole moment orientation, the average $C_\mathrm{fTE}$ for axial defects remains the same; however, for the basal defects, it reduces to $\cos(71^{\circ})\cdot C_\mathrm{fTE}$ = 6.55\%. Assuming a 1:1 ratio for axial and basal, the overall $C_\mathrm{fTE}$ value becomes 13.32\%. 

\section{COLLECTION EFFICIENCY CALCULATION}
From the 1D PL scans (in the main text), we obtain the normalized PL intensity from the top of a 10 $\mu$m long triangular waveguide (where the NV centers are excited) and the periodic and aperiodic GCs. As the NV center emission spectrum is close to 1240 nm, for the efficiency calculations, we utilize the waveguide loss of 0.3 dB/$\mu$m at 1240 nm, as presented in Table I (main text). As we excite the center of the waveguide, the PL emission travels 5 $\mu$m before reaching the GC. Hence, the total propagation loss amounts to 1.5 dB, corresponding to a transmission efficiency of 70.8\%. 

Considering the calculated average 13.32\% coupling of NV ensemble emission to the f-TE mode of the waveguide and the waveguide transmission efficiency of 70.8\%, the accumulated system efficiency reduces to 9.4\%. In both periodic and aperiodic configurations, the normalized maximum PL (1 a.u.) is collected from the top surface at the waveguide center. Considering the $C_\mathrm{wvg, top}$ value to be 11.29\%, the actual normalized PL emission in the waveguide becomes 8.857 a.u., and multiplying the value by the system efficiency of 9.4\%, we obtain the normalized PL emission of 0.8326 a.u. reaching the GC end. As we observe a slightly asymmetric collection profile from the two GC ends, we report both values for periodic and aperiodic designs in the following table.

\begin{table}[H]
\caption{\label{tab:gc_efficiency} Grating coupler efficiency}
\centering
\resizebox{\textwidth}{!}{%
\begin{tabular}{|c|c|c|c|c|}
\hline
\textbf{GC} & \textbf{Position} & \textbf{Norm. PL before GC (a.u.)} & \textbf{Norm. PL after GC (a.u.)} & \textbf{Collection efficiency (\%)} \\
\hline
Periodic & Right & 0.8326 & 0.15 & 18 \\
Periodic & Left & 0.8326 & 0.19 & 23 \\
Aperiodic & Right & 0.8326 & 0.175 & 21 \\
Aperiodic & Left & 0.8326 & 0.20 & 24 \\
\hline
\end{tabular}%
}
\vspace{1ex}
\end{table}

\clearpage

\bibliographystyle{unsrt}
\bibliography{bib}
\end{document}